\documentclass[final,1p,times]{elsarticle} 

\usepackage{graphicx}
\usepackage{amssymb} 
\usepackage{amsthm} 
\usepackage{lineno}

\journal{Nuclear Physics A} 

\begin{document}

\begin{frontmatter} 

\title{Lattice calculation of the heavy quark potential at non-zero temperature}

\author[auth1]{A. Bazavov} 
\author[auth2]{Y. Burnier}
\author[auth3]{P. Petreczky}
\address[auth1]{Department of Physics and Astronomy, University of Iowa,
203 Van Allen Hall, Iowa City, Iowa 52242-1479 }
\address[auth2]{Laboratory of Particle Physics and Cosmology, \'Ecole polytechnique f\'ed\'erale 
de Lausanne, BSP 730,
Rte de la Sorge,
CH-1015 Lausanne}
\address[auth3]{Physics Department, Brookhaven National Laboratory, Upton, NY 11793, USA}

\begin{abstract} 
We calculated the real and imaginary parts of the static quark anti-quark
potential at $T>0$ in 2+1 flavor QCD using correlators of Wilson lines
in Coulomb gauge and lattices with temporal extent $N_{\tau}=12$. We find
that the real part of the potential is larger than the singlet free energy
but smaller than the zero temperature potential. The imaginary part of the
potential is similar in size to the perturbative HTL result.
\end{abstract} 

\end{frontmatter} 


\section{Introduction}

There has been significant interest in studying quarkonium properties
at non-zero temperature since the seminal paper by Matsui and Satz \cite{MS86}
(for a recent review see e.g. Ref. \cite{merev}).
It has been conjectured that the thermal medium will modify the heavy quark potential, 
eventually leading to the dissolution of the heavy quarkonium states \cite{MS86}.
The problem of dissolution of
quarkonium states at high temperatures can be rigorously formulated in terms of spectral
functions. Early attempts to calculate the quarkonium spectral functions on the lattice
have been presented in Refs. \cite{latspf}. However, extraction of the spectral functions
from lattice results on Euclidean correlation functions is quite difficult \cite{ines01}
and one should also be careful with cutoff effects in the spectral functions extracted from
the lattice \cite{freelatspf}. Furthermore it is difficult to observe the melting of bound states at high temperature in Euclidean correlators due to the fact that their time extent is limited to $<1/(2T)$ (see e.g. discussions in Ref. \cite{merev,merev1})
and alternative method to determine the spectral functions could be useful.

The effective field theory framework for heavy quark bound states, namely pQNRCD could be
a useful tool for calculating quarkonium spectral functions \cite{nora}. 
The effective field theory approach allows a rigorous definition of  the concept of the 
static quark anti-quark potential both at zero and non-zero temperatures. One of the main
outcomes of the effective field theory analysis is the finding that at non-zero temperature
the potential has also an imaginary part, which has important consequences for the
dissolution of the quarkonium states. While pNRQCD is 
formulated in the weak coupling framework it is possible to extend it to the non-perturbative 
regime. For example, if the binding energy is the smallest scale in the problem all
the other scales, like the thermal scales, the inverse size of the bound state and $\Lambda_{QCD}$
can be integrated out. In this case the potential is identical to the energy of a static $Q\bar Q$ pair
and can be calculated non-perturbatively on the lattice.
Therefore, in what follows we will use the terms potential and static energy interchangeably.
If one further neglects the dipole interactions one gets the generalization of the
simple potential model to the case of high temperatures \cite{me10}. 
In Ref. \cite{rothkopf} it has been suggested
to extract the energy of a static $Q\bar Q$ pair using the spectral decomposition of the 
temporal Wilson loops at non-zero temperature. In this contribution we calculate the 
quark anti-quark energy in 2+1 flavor QCD based on this idea using lattices
with temporal extent $N_{\tau}=12$. Previously we calculated the real part of the static energy 
using $N_{\tau}=6$ lattices \cite{qm12}.

\section{Numerical results}
\begin{figure} 
\includegraphics[width=7cm]{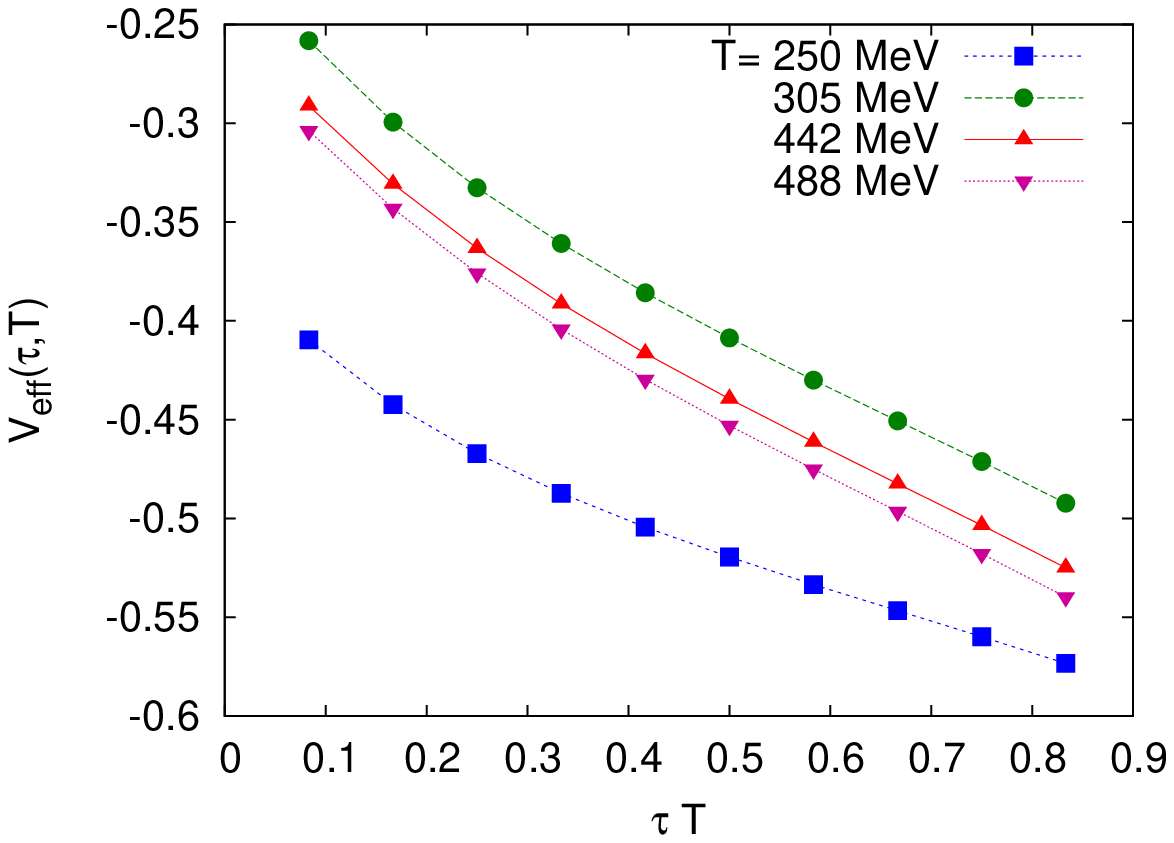} 
\includegraphics[width=7cm]{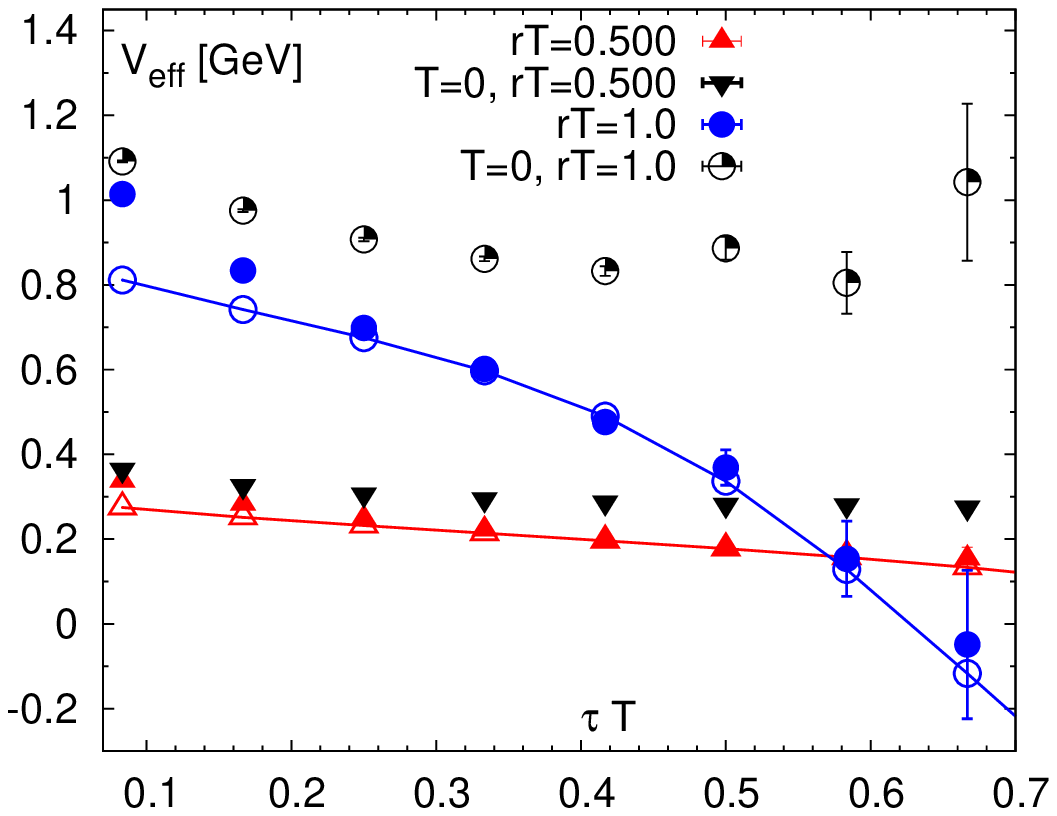} 
\caption{The effective potential as function of $\tau$ in HTL perturbation theory (left) and in lattice calculations (right).
The line-points in the right panel represent the fit to the lattice data based on HTL Ansatz (see text). }
\end{figure}
In lattice QCD calculations the static $Q\bar Q$ energy is extracted from Wilson loops
$W(r,\tau)$. At large Euclidean time separations the exponential decay of the
Wilson loops is governed by the static energy or potential, $W(r,\tau)\sim \exp(-V(r)\tau)$.
More generally one can write a spectral decomposition for the Wilson loops \cite{rothkopf}
\begin{equation}
W(r,\tau)=\int_{-\infty}^{\infty} d \omega \sigma(r,\omega) e^{-\omega \tau}.
\label{spectral}
\end{equation}
At zero temperature the spectral function is proportional to $\delta(\omega-V(r))$ at sufficiently low $\omega$. At non-zero temperature the delta
function becomes a skewed Lorentzian with the width related to the imaginary part of the potential \cite{Burnier:2012az}.
Lattice calculations of the potential $T>0$ using Eq. (\ref{spectral}) and maximum entropy method (MEM)
in SU(3) gauge theory have been presented in Ref. \cite{rothkopf}.
We would like to calculate the potential in 2+1 flavor QCD.

We use gauge configurations in 2+1 flavor QCD generated on $48^3 \times 12$ lattices using highly
improved staggered quark (HISQ) action \cite{follana} using physical quark strange quark mass and light
quark masses corresponding to pion mass of $160$ MeV in the continuum limit. 
We consider two values of the gauge coupling $\beta=10/g^2=7.28$ and $7.50$ corresponding to temperatures
$T=250$ MeV and $305$ MeV. The choice of the lattice parameters is discussed in Ref. \cite{tc}. 
There are two challenges when one tries to extract the potential using $N_{\tau}=12$ lattices. First,
it is not possible to obtain reliable results using MEM on $N_{\tau}=12$ lattices. The second problem
is the poor overlap of simple (unsmeared) Wilson loops with the ground state $Q\bar Q$ energy.  
To deal with the second problem smeared gauge
fields are used in spatial links when constructing Wilson loops on the lattice. Alternatively,
one can fix the Coulomb gauge and calculate the correlation functions of two temporal Wilson
lines separated by distance $r$ without connecting them by spatial links \cite{milc04}.
Both method have been used in the past at zero temperature and very recently also at $T>0$ \cite{epja}.
The $\tau$ dependence of the smeared Wilson loop and Wilson line correlators is very similar \cite{epja},
and therefore, in this study we will use Wilson line in Coulomb gauge.
\begin{figure}
\includegraphics[width=7cm]{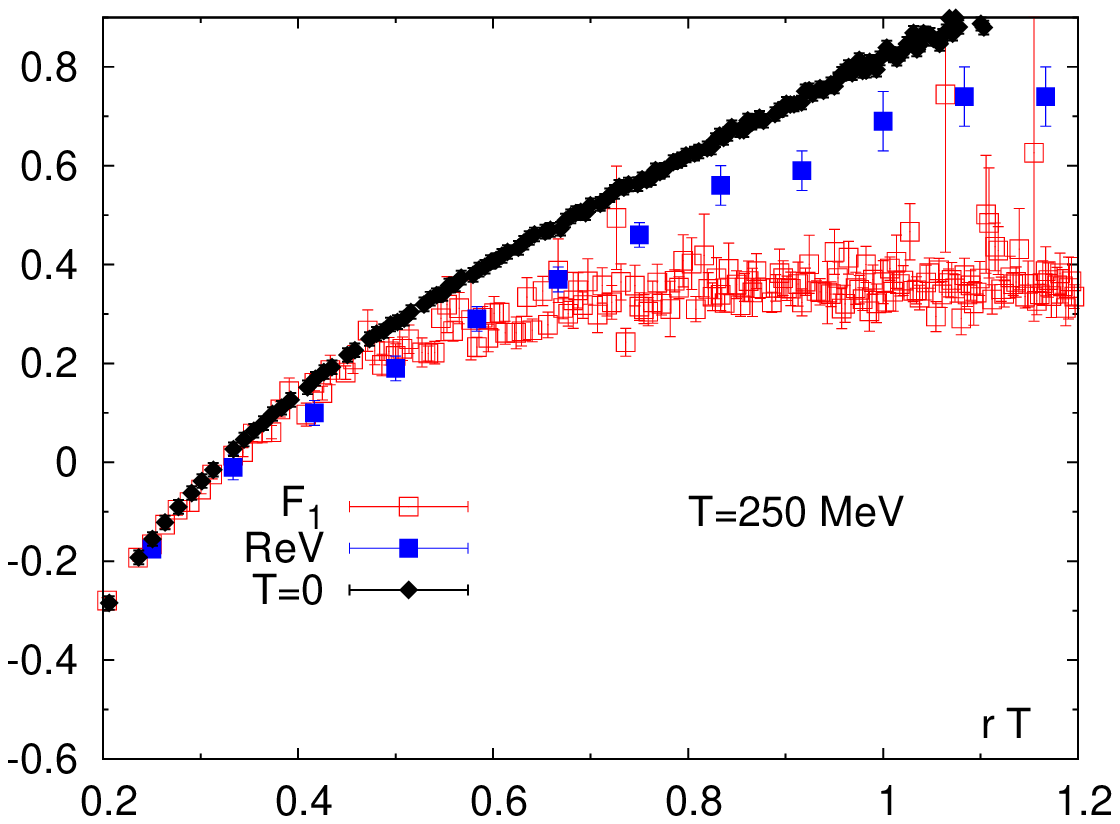}
\includegraphics[width=7cm]{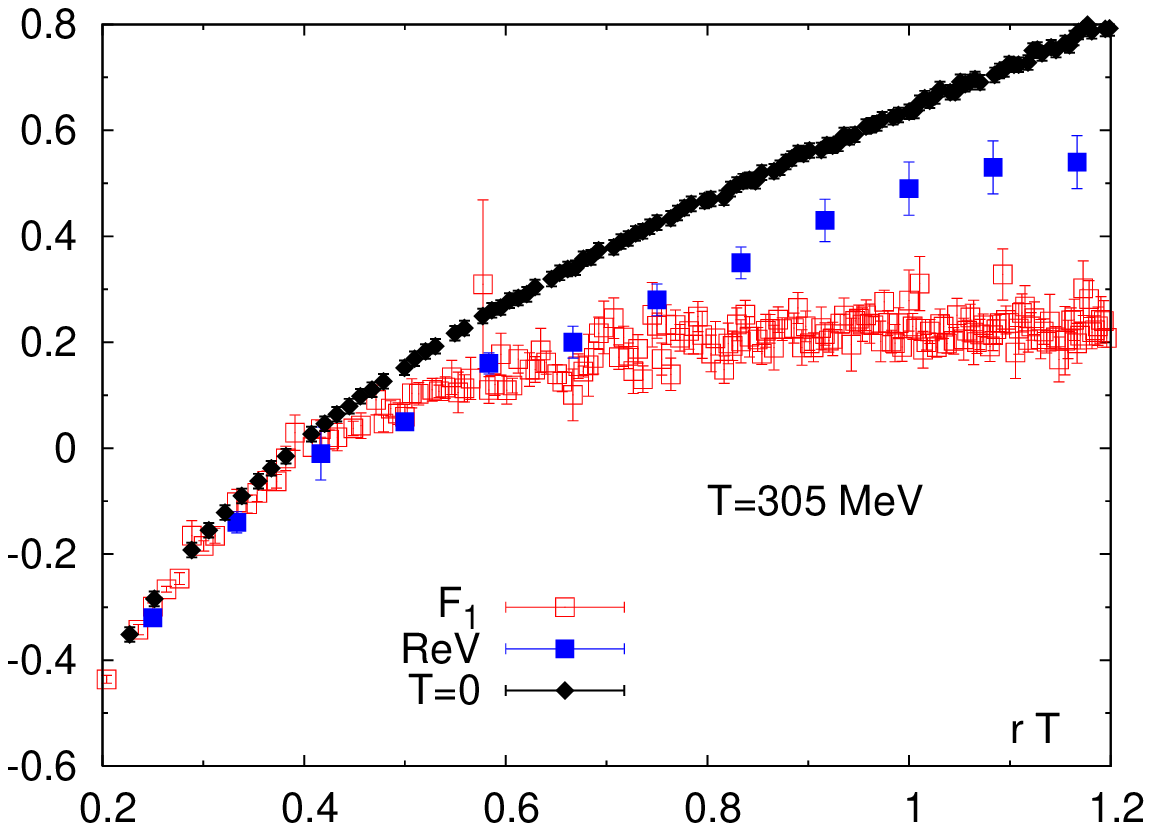}
\caption{The real part of the potential at $T=250$ MeV and $T=305$ MeV compared to
the zero temperature potential and the singlet free energy.}
\end{figure}
To deal with the first problem we will make use of the analytic calculations of the spectral
function of Wilson line correlators in Coulomb gauge in Hard Thermal Loop (HTL) perturbation
theory \cite{htl}. In Ref. \cite{htl} the spectral functions corresponding to Wilson loops
and correlators of Wilson lines in Coulomb gauge has been calculated. For the Wilson loops
the spectral function is quite complicated and the ground state peak is not the dominant structure.
In the case of the Wilson line correlators the spectral function is very simple, it consist of a 
single peak, though the peak is asymmetric and has long tails \cite{htl}. 
The simple structure of the spectral functions makes the correlators of Wilson lines in Coulomb gauge 
suitable for extraction of the potential at $T>0$. 

It is useful to analyze the
$\tau$ -dependence of this correlator in terms of the effective potential
$V_{\rm eff}(r,\tau)=-\ln(W(r,\tau)/\tau)$. For $T=0$, $V_{\rm eff}(r,\tau)$ reaches a plateau
at sufficiently large $\tau$ which is given by the static $Q\bar Q$ potential. In Fig. 1 we show
the $V_{\rm eff}(r,\tau)$ at zero and finite temperature in lattice calculations. We see that at
zero temperature the plateau is reached already for $\tau T \simeq 0.5$, but for $T>0$ the effective potential
always decreases without reaching a plateau. This is due to the width of the peak, i.e. to
the imaginary part of the potential. We calculated the effective potential using
the results on spectral functions in HTL perturbation theory \cite{htl} which show the same 
qualitative $\tau$ dependence as observed in lattice calculations. Therefore, we extract the real
and imaginary part of the potential by fitting the $\tau$-dependence 
of the lattice results on $V_{\rm eff}(r,\tau)$ by the two parameter Ansatz $\sigma^{HTL}(\lambda (\omega-E))$ with $E$ 
and $\lambda$ being the fit parameters
that determined the peak position and its width. The fits are shown are also shown in Fig. 1 as line-points.
We do not attempt to fit the points and small $\tau$ as the HTL Ansatz is not expected to work there.
\begin{figure}
\includegraphics[width=7cm]{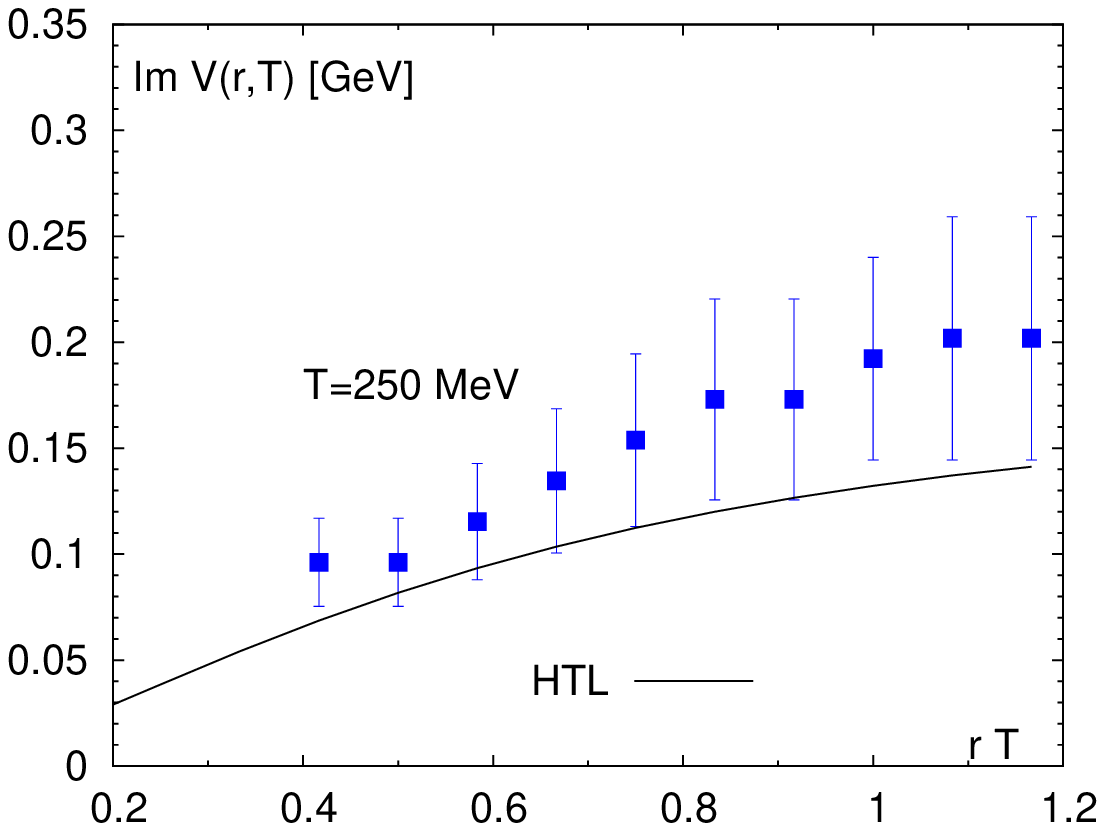}
\includegraphics[width=7cm]{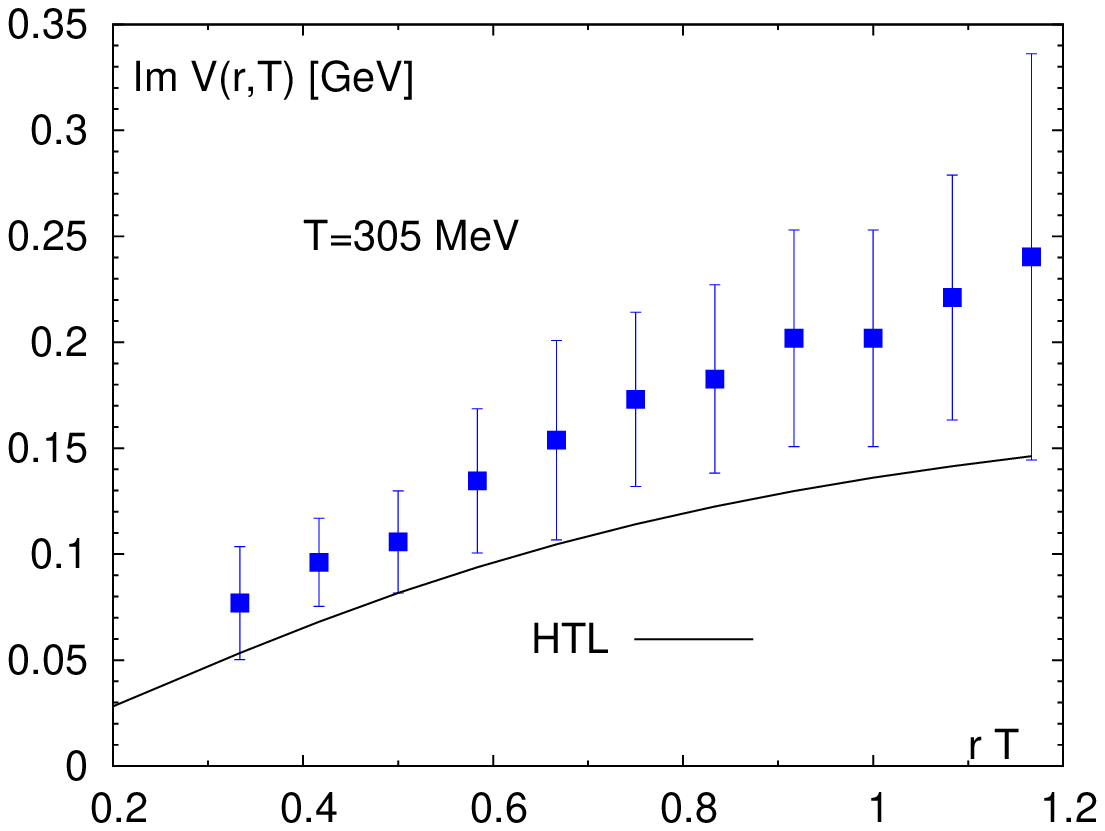}
\caption{The imaginary part of the potential at $T=250$ MeV and $T=305$ MeV compared to the HTL result \cite{laine06}.}
\end{figure}
The results of this analysis are are shown in Figs. 2 and 3. The errors shown in the figure are mostly systematic
and arise from the variation of the fit interval. We compared the real part of the potential with the zero
temperature result as well as with the singlet free energy. The real part of the potential is equal or
larger than the singlet free energy but is always smaller than the zero temperature potential. Furthermore,
it decreases with increasing temperature in qualitative agreement with the picture of color screening.
The central value of the imaginary part of the potential turns out to be larger than the imaginary
part of the potential in HTL perturbation theory though withing errors it is the same.

\section{Conclusion}
We have extracted the real and imaginary parts of the potential using lattice results
on the Wilson line correlators in Coulomb gauge on $N_{\tau}=12$ lattices and a fit
Ansatz motivated by HTL perturbation theory. Our results on the real part of the 
potential are in agreement with our previous findings obtained using $N_{\tau}=6$ lattices
and more simplistic ad-hoc form for the corresponding spectral function \cite{qm12}.
We were also able to obtain the imaginary part of the potential which turns out to be larger
than in HTL perturbation theory.

\noindent
{\bf Acknowledgment:}
This work was partly supported by through the Contract No. DE-AC02-98CH10886 with the U.S. Department of Energy.
YB was supported by the SNF under grant PZ00P2-142524.

\end{document}